\documentclass[journal]{IEEEtran}

\usepackage{amsmath,amssymb}
\usepackage{graphicx}
\usepackage{cite}
\usepackage{algorithm}
\usepackage{algpseudocode}
\usepackage{eso-pic}

\begin{document}

\AddToShipoutPictureFG*{%
  \AtPageLowerLeft{%
    \raisebox{7mm}{%
      \makebox[\paperwidth][c]{%
        \parbox{0.92\paperwidth}{%
          \centering\footnotesize
          This work has been submitted to the IEEE for possible publication.
          Copyright may be transferred without notice, after which this
          version may no longer be accessible.
        }%
      }%
    }%
  }%
}
\title{Causality-Aware Interaction Selection for Retarded Green-Function Assembly}

\author{%
Sushil~Kumar,
Wolf~van~der~Hert,
Giampiero~Gerini,~\IEEEmembership{Senior Member,~IEEE},
and M.~C.~van~Beurden,~\IEEEmembership{Senior Member,~IEEE}%
\thanks{S. Kumar, W. van der Hert, G. Gerini, and M. C. van Beurden
are with the Department of Electrical Engineering, Eindhoven University
of Technology, Eindhoven, The Netherlands
(e-mail: s.kumar2@tue.nl; m.c.v.beurden@tue.nl).
Corresponding author: S. Kumar.}%
\thanks{G. Gerini is also with the Optics Department, Netherlands
Organisation for Applied Scientific Research (TNO), Delft,
The Netherlands (e-mail: giampiero.gerini@tno.nl).}%
}


\maketitle

\begin{abstract}
In time-domain integral-equation (TDIE) solvers, the assembly of the retarded Green-function interactions remains a major computational bottleneck. We present a causality-aware assembly strategy that exploits the finite space--time support of the retarded Green function to identify admissible interactions prior to numerical evaluation. By using causality as an exact interaction-selection criterion, causally inadmissible interactions are excluded \emph{a priori}, reducing assembly complexity without modifying the underlying TDIE formulation or introducing approximations. Numerical validation confirms that the proposed pruning preserves the TDIE transient response. The remaining admissible interactions are organized into vectorized shared-memory workloads for efficient assembly. The results demonstrate a 41\% reduction in evaluated interactions, a 2.6 times single-worker algorithmic speedup, and a total assembly-time speedup of up to 109 times using 64 CPU workers.
\end{abstract}

\begin{IEEEkeywords}
Causality, Green-function assembly, marching-on-in-time, retarded potential, time-domain integral equations, transient electromagnetic scattering, volume integral equations.
\end{IEEEkeywords}

\section{Introduction}
Broadband electromagnetic analysis of dielectric scatterers, resonators, nanophotonic structures, and metasurfaces requires an accurate treatment of radiation and wave propagation in open regions. Time-domain integral-equation (TDIE) methods are attractive for such problems because the retarded Green function naturally satisfies the radiation conditions and enables broadband responses from a single transient simulation \cite{Shanker2009_TDIE_CompositeBodies,Ren2022_TDCEM}.  A major computational bottleneck in practical TDIE solvers is the assembly of the retarded Green-function interactions over source and observation regions, across temporal delay indices. In marching-on-in-time (MOT) formulations, traditional assembly procedures evaluate a large number of such candidate interactions determined by retarded delays and temporal basis supports. Although significant advances have been made in TDIE stability, fast convolution, FFT-based acceleration, and parallel implementations \cite{VantWout2013_TemporalBasis,Sayed2015_TD_EFVIE_HighContrast,VanDiepen2024_MOTJVIE_FFT}, these approaches primarily reduce the cost of evaluating interaction operators. Many candidate interactions, however, are inactive because they lie outside the causal space--time support of the retarded Green function. 

We present a causality-aware Green-function assembly strategy for TDIE solvers. The proposed method converts retarded causality into an exact pre-assembly admissibility test that identifies active source--observation--delay interactions before numerical evaluation. Interactions outside the causal support are excluded a priori, reducing assembly complexity without modifying the underlying TDIE formulation or introducing approximations. The remaining admissible interactions are organized into vectorization-friendly shared-memory parallel workloads for efficient evaluation. The approach is demonstrated using a causal implicit MOT-JVIE solver \cite{VanDiepen2024_MOTJVIE_Contrast} for dielectric scattering. Numerical results verify preservation of the transient scattered-field response and quantify interaction reduction, assembly-time acceleration, memory usage, and parallel scalability.

\section{Time-Domain MOT-JVIE Formulation}

The proposed assembly strategy is demonstrated for the causal implicit marching-on-in-time current-density volume integral equation (MOT-JVIE) in  \cite{VanDiepen2024_MOTJVIE_Contrast}. For a dielectric object occupying volume \(V_{\varepsilon}\), the time-domain current-density volume integral equation is
\begin{multline}
\left(
\varepsilon_r(\mathbf r)-1
\right)
\varepsilon_0
\frac{\partial}{\partial t}
\mathbf E^{i}(\mathbf r,t)
=
\varepsilon_r(\mathbf r)
\mathbf J_{\varepsilon}(\mathbf r,t)
\\
-
\left(
\varepsilon_r(\mathbf r)-1
\right)
\nabla\times\nabla\times
\iiint_{V_{\varepsilon}}
\frac{
\mathbf J_{\varepsilon}(\mathbf r',\tau_0)
}{
4\pi R
}
\,dV',
\label{eq:TDJVIE}
\end{multline}
where
\begin{equation}
\mathbf J_{\varepsilon}(\mathbf r,t)
=
\left(
\varepsilon(\mathbf r)-\varepsilon_0
\right)
\frac{\partial \mathbf E(\mathbf r,t)}
{\partial t},
\label{eq:Contrast_current_density}
\end{equation}
with \(R=\|\mathbf r-\mathbf r'\|\), \(\tau_0=t-R/c_0\),
and \(\varepsilon_r=\varepsilon/\varepsilon_0\), where
\(\mathbf r\) and \(\mathbf r'\) denote the observation and
source coordinates, respectively. After space--time discretization \cite{VanDiepen2024_MOTJVIE_Contrast}, the causal implicit MOT system becomes
\begin{equation}
\mathbf Z_0\mathbf J_n
=
\mathbf E_n^{i}
-
\sum_{n'=n-1}^{n-\ell}
\mathbf Z_{n-n'}
\mathbf J_{n'},
\label{eq:mot_tdjvie}
\end{equation}
where \(\mathbf J_n\) is the unknown contrast-current vector at time step \(n\), and \(\mathbf Z_{n-n'}\) contains the retarded Green-function interactions associated with delay index \(n-n'\). The maximum delay level is
\begin{equation}
\ell_{\max}
=
\left\lceil
\frac{R_{\max}}
{c_0\Delta t}
\right\rceil
+ p,
\label{eq:lmax}
\end{equation}
where \(R_{\max}\) denotes the maximum observation--source separation distance, \(\Delta t\) is the time-step size, and \(p\) is temporal-basis order.

\section{Causality-Aware Green-Function Assembly}

\subsection{Causality-Aware Interaction Selection}
\label{subsec:causal_interaction_selection}

Although the MOT history is bounded by \(\ell_{\max}\), each observation--source voxel pair occupies only a subset of this history \cite{Weile2004_TDIE_NovelScheme,Shanker2000_TDCFIE_ClosedSurfaces}. The retarded delay between observation voxel \(m\) and source voxel \(m'\) is \(\tau_{mm'}=R_{mm'}/c_0\), where \(R_{mm'}\) is the observation--source separation distance \cite{Sayas2016_RetardedPotentials}.

\begin{figure}[htb]
\centering
\includegraphics[width=0.65\columnwidth]{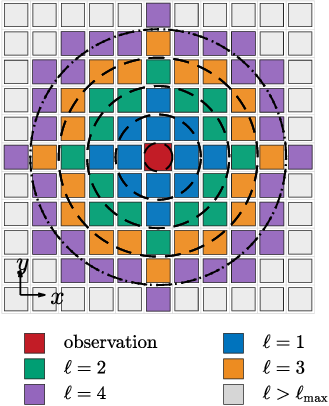}
\caption{Causality-aware organization of retarded interactions. For a fixed observation voxel \(m\), source voxels \(m'\) are grouped into discrete delay shells according to the retarded delay \(\tau_{mm'}=R_{mm'}/c_0\).}
\label{fig:causality_aware_concept}
\end{figure}

As shown in Figure~\ref{fig:causality_aware_concept}, this delay maps observation and source voxel pairs into discrete delay shells around the observation voxel. A source voxel can contribute only when its retarded delay overlaps the support of the temporal basis function \(T(t)\) used in the MOT expansion \cite{Sauter2013_RetardedBIE_CompactTemporalBasis,Weile2004_TDIE_NovelScheme}. Interactions outside this causal-support region are therefore identically inactive and need not be assembled. Defining the MOT delay index \(k=n-n'\), a observation--source interaction is admissible only if
\begin{equation}
k\Delta t-\tau_{mm'}\in[t_a,t_b],
\label{eq:basis_overlap}
\end{equation}
where \([t_a,t_b]\) denotes the support of \(T(t)\). Equation~(\ref{eq:basis_overlap}) states that an observation--source interaction is admissible only when the retarded delay overlaps the support of the temporal basis function. Solving (\ref{eq:basis_overlap}) for the delay index \(k\) yields the admissible delay-index set

\begin{equation}
\mathcal{Q}_{mm'}
=
\left\{
k:
\left\lceil
\frac{\tau_{mm'}+t_a}{\Delta t}
\right\rceil
\le
k
\le
\left\lfloor
\frac{\tau_{mm'}+t_b}{\Delta t}
\right\rfloor
\right\},
\label{eq:active_delay_set}
\end{equation}
which depends only on the observation--source separation through \(\tau_{mm'}\) and the temporal-basis support. The corresponding admissible space--time interaction set is

\begin{equation}
\begin{aligned}
\mathcal{S}_{\mathrm{act}}
=
\{(m,m',k):\;&1\le m,m'\le N_v,\;
0\le k\le \ell_{\max},\\
&k\in\mathcal{Q}_{mm'}\},
\end{aligned}
\label{eq:active_set}
\end{equation}
where \(N_{v}\) is the total number of voxels. All remaining candidate interactions are removed before the elements in the matrices pertaining to the  Green function are evaluated, because the retarded delay and temporal-basis support do not overlap. Consequently, the proposed pruning preserves the original TDIE formulation and introduces no approximation, as only causally inadmissible space--time interactions are excluded. The resulting admissible interaction set \(\mathcal{S}_{\mathrm{act}}\) is organized into vectorization-friendly shared-memory parallel workloads for efficient Green-function assembly. Hence, causality is exploited as an exact pre-assembly interaction-selection criterion, reducing the number of assembled interactions and the associated assembly complexity.

\subsection{Vectorized CPU-Parallel Assembly}
\label{subsec:vectorized_batching}

After causality-aware interaction selection, the admissible interaction set \(\mathcal{S}_{\mathrm{act}}\) contains all observation--source--delay triples \((m,m',k)\). As discussed in Section~\ref{subsec:causal_interaction_selection}, interactions are retained in \(\mathcal{S}_{\mathrm{act}}\) only when their retarded-delay range overlaps \(k\Delta t\); otherwise, they are discarded before quadrature. This avoids evaluating Green-function interactions whose individual contributions are guaranteed to vanish under the retarded causality condition, thereby eliminating repeated light-cone conditional checks. By analyzing the statistics of the computation time over the various tasks, we came to the conclusion that a significant portion of the CPU time is spent on branching between the different cases for the pertaining integrals. Therefore, the retained interactions are subsequently grouped into homogeneous batches such that interactions following the same evaluation path are processed together using branch-free vectorized array operations. Each admissible interaction contributes to a retarded Green-function entry \(\mathbf{Z}_{mm'}^{(k)}\), which is accumulated in the corresponding delay-indexed interaction matrix \(\mathbf{Z}_{k}\)~\cite{Weile2004_TDIE_NovelScheme}.

The homogeneous interaction batches \(\{\mathcal{B}_{b}\}_{b=1}^{N_{\mathrm{batch}}}\), where \(\mathcal{B}_{b}\subset\mathcal{S}_{\mathrm{act}}\), are distributed among \(P\) CPU workers to improve memory locality and enable single-instruction, multiple-data (SIMD)~\cite{openmp52} aware shared-memory execution~\cite{Markall2013_FEMAssembly_Multicore,Trott2022_Kokkos3}. To avoid synchronization overhead, worker \(r\) accumulates its assigned interaction contributions into private delay-indexed buffers \(\{\mathbf{Z}_{r,k}\}_{k=0}^{\ell_{\max}}\). After all batches have been processed, the global interaction matrices are obtained as
\begin{equation}
\mathbf{Z}_{k}
=
\sum_{r=1}^{P}
\mathbf{Z}_{r,k},
\qquad
k=0,\ldots,\ell_{\max}.
\end{equation}

The batching improves memory locality and enables efficient vectorized evaluation of retarded Green-function interactions \cite{Luporini2015_CrossLoopAssembly,Trott2022_Kokkos3}. Combined with the causality-aware interaction-selection procedure, this strategy substantially reduces the computational cost of constructing the retarded Green-function interaction history.

\section{Numerical Validation and Performance}

Electromagnetic scattering from two dielectric slabs under visible-wavelength illumination is considered to assess the proposed causality-aware assembly strategy. The two configurations are referred to as the small- and large-slab cases. The slab material is modeled as a lossless, non-dispersive dielectric with relative permittivity \(\varepsilon_r=12\). The structure is illuminated by an \(x\)-polarized modulated Gaussian plane wave propagating along \(-\hat{\mathbf z}\),

\begin{equation}
\begin{aligned}
\mathbf{E}^{\mathrm{inc}}(\mathbf r,t)
&=
E_0\hat{\mathbf x}
\exp\!\left(
-\frac{\tau^2}{2\sigma^2}
\right)
\cos\!\left(
2\pi f_0\tau
\right), \\
\tau
&=
t-t_0-\frac{\hat{\mathbf k}\cdot\mathbf r}{c_0},
\qquad
\hat{\mathbf k}=-\hat{\mathbf z}.
\end{aligned}
\label{eq:Einc_modGauss}
\end{equation}

The incident-field parameters is summarized in Table~\ref{tab:incident_parameters}. The slab dimensions, spatial discretization, temporal discretization are summarized in Table~\ref{tab:discretization_parameters}. The small slab case is used to verify that the proposed causality-aware assembly preserves the transient response obtained using the conventional assembly. The large slab case is used to quantify interaction reduction, assembly performance, and parallel scalability. Owing to its larger physical dimensions and correspondingly longer propagation delays, the large slab case employs a longer simulation time ($T_{\mathrm{sim}}$) and a larger maximum delay level $\ell_{\max}$.

\begin{table}[htb]
\centering
\footnotesize
\caption{Incident-field parameters for the dielectric slabs}
\label{tab:incident_parameters}
\setlength{\tabcolsep}{2pt}
\renewcommand{\arraystretch}{0.92}
\begin{tabular}{lccc}
\hline
\hline
\textbf{Parameter} & \textbf{Description} & \textbf{Value} & \textbf{Units} \\
\hline
\(E_0\) & Peak field amplitude & 0.02 & \(\mathrm{V/m}\) \\
\(f_0\) & Center frequency & 793.5 & \(\mathrm{THz}\) \\
\(\lambda_0\) & Center wavelength & 378 & \(\mathrm{nm}\) \\
\(\sigma\) & Pulse width & 0.40 & \(\mathrm{fs}\) \\
\(t_0\) & Pulse delay & 2.52 & \(\mathrm{fs}\) \\
\hline
\end{tabular}
\end{table}

 For both slab cases, the conventional and proposed assemblies employ identical slab geometries, voxel discretizations, temporal basis functions, time steps, excitation parameters, material properties, and MOT-JVIE settings; only the Green-function assembly procedure differs. Consequently, all observed assembly-time reductions arise from reducing the number of assembled retarded Green-function interactions through causal-support screening and from the efficient vectorized parallel assembly of the resulting admissible interactions.

\begin{table}[htb]
\centering
\footnotesize
\caption{Simulation parameters for the dielectric slabs}
\label{tab:discretization_parameters}
\setlength{\tabcolsep}{2pt}
\renewcommand{\arraystretch}{0.92}
\begin{tabular}{lcc}
\hline
\hline
\textbf{Parameter} & \textbf{Small slab} & \textbf{Large slab} \\
\hline
Slab size \((\mathrm{nm}^3)\) & \(400\times400\times200\) & \(2000\times2000\times200\) \\
Grid, \(N_x\times N_y\times N_z\) & \(40\times40\times20\) & \(200\times200\times20\) \\
Voxel size \((\mathrm{nm}^3)\) & \(10\times10\times10\) & \(10\times10\times10\) \\
No. of voxels, \(N_v\) & \(3.2\times10^4\) & \(8.0\times10^5\) \\
Temporal basis order \((p)\) & 2 & 2 \\
Time step, \(\Delta t\) & \(0.02~\mathrm{fs}\) & \(0.02~\mathrm{fs}\) \\
Total simulation time, \(T_{\mathrm{sim}}\) & \(8~\mathrm{fs}\) & \(40~\mathrm{fs}\) \\
Number of time steps, \(N_t\) & \(400\) & \(2000\)\\
\hline
\end{tabular}
\end{table}

\subsection{Transient-Field Accuracy Validation}
\begin{figure}[htb]
\centering
\includegraphics[width= 1\columnwidth]{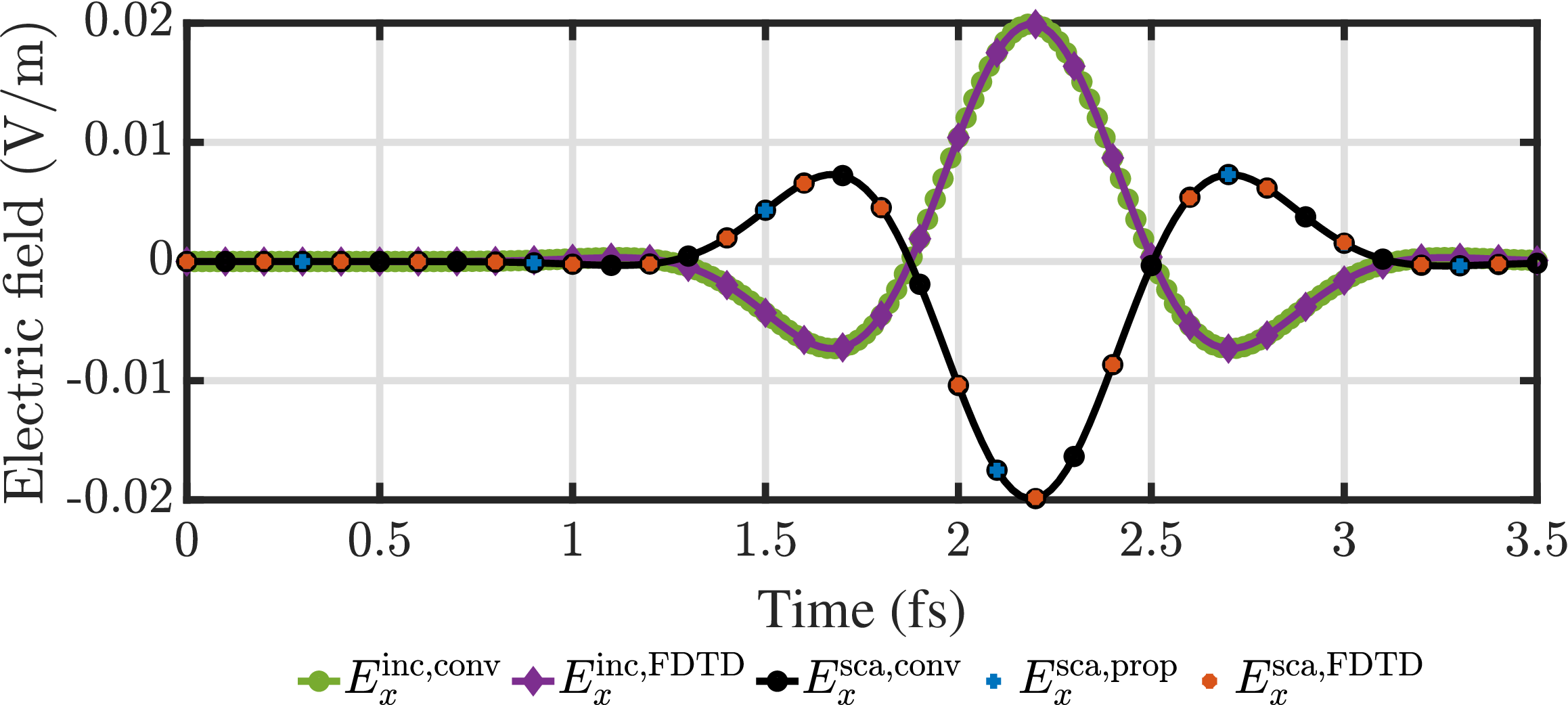}
\caption{Small-slab accuracy validation: transient \(x\)-directed scattered fields at the slab center from conventional MOT-JVIE, causality-aware MOT-JVIE, and Meep FDTD, with incident fields verifying consistent excitation.}
\label{fig:validation_slab}
\end{figure}

The small slab in Table~\ref{tab:discretization_parameters} is used to verify that the proposed causality-aware assembly preserves the MOT-JVIE transient response. The conventional assembly uses retarded Green-function assembly as used in \cite{VanDiepen2024_MOTJVIE_Contrast}, whereas the proposed assembly retains only source--observation--delay interactions satisfying the causal-support criterion governed by Equation~\eqref{eq:active_set}. An independent three-dimensional finite-difference time-domain (FDTD) validation was also performed using Meep~\cite{Oskooi2010_Meep}, where the scattered field $\mathbf{E}_x^{\mathrm{sca,FDTD}}$ was obtained from the difference between the slab total-field response and the empty-cell incident-field response $\mathbf{E}_x^{\mathrm{inc,FDTD}}$. The dielectric slab occupies the rectangular domain
\(0\le x\le400~\mathrm{nm}\),
\(0\le y\le400~\mathrm{nm}\), and
\(0\le z\le200~\mathrm{nm}\), with its faces aligned with the Cartesian coordinate axes. The \(x\)-directed scattered electric field is recorded at the slab center,
\(\mathbf r_{\mathrm{obs}}=(200,200,100)~\mathrm{nm}\),
as shown in Fig.~\ref{fig:validation_slab}. The agreement between the proposed and conventional assemblies is quantified using the normalized waveform error

\begin{equation}
\epsilon_\mathbf{E}
=
\frac{
\left(
\sum_{n=1}^{N_t}
\left|
\mathbf{E}_{x,n}^{\mathrm{sca,prop}}
-
\mathbf{E}_{x,n}^{\mathrm{sca,conv}}
\right|^2
\right)^{1/2}
}{
\left(
\sum_{n=1}^{N_t}
\left|
\mathbf{E}_{x,n}^{\mathrm{sca,conv}}
\right|^2
\right)^{1/2}
},
\label{eq:relative_error}
\end{equation}

where \(N_t\) is the total number of sampled time steps, and
\(\mathbf{E}_{x,n}^{\mathrm{sca,prop}}\) and
\(\mathbf{E}_{x,n}^{\mathrm{sca,conv}}\) denote the \(x\)-directed scattered-field samples obtained using the proposed and conventional assemblies, respectively.

The agreement in Figure~\ref{fig:validation_slab}, with \(\epsilon_\mathbf{E}\simeq10^{-15}\), confirms that the causality-aware assembly preserves the conventional MOT-JVIE transient scattered-field response. The FDTD incident and scattered-field responses also agree well with the corresponding conventional MOT-JVIE results, providing an independent verification of the excitation and scattered-field extraction procedures.

\subsection{Interaction Reduction and CPU-Parallel Performance}
The large slab in Table~\ref{tab:discretization_parameters} is used to evaluate the computational benefit of the proposed causality-aware assembly. Before numerical evaluation of the Green function, the causal-support criterion is applied to each candidate entry of the retarded Green function table. Because the voxel grid is uniform, the implementation exploits shift symmetry. 

\begin{table}[htb]
\caption{Causality-aware Green-function interaction statistics.}
\label{tab:causality_statistics}
\centering
\footnotesize
\setlength{\tabcolsep}{4pt}
\renewcommand{\arraystretch}{0.95}
\begin{tabular}{lc}
\hline
\hline
\multicolumn{2}{c}{\textbf{(a) Overall interaction statistics}} \\
\hline
\textbf{Metric} & \textbf{Value} \\
\hline
Total candidate entries, \(N_{\rm tot}\) & \(120{,}408{,}556\) \\
Active entries, \(N_{\rm act}\) & \(71{,}468{,}436\) \\
Skipped entries, \(N_{\rm skip}\) & \(48{,}940{,}120\) \\
Active fraction, \(\eta_{\rm act}\) & \(59.4\%\) \\
Skipped fraction, \(\eta_{\rm skip}\) & \(40.6\%\) \\
\end{tabular}
\begin{tabular}{ccccc}
\hline
\hline
\multicolumn{5}{c}{\textbf{(b) Per-level interaction statistics}} \\
\hline
\textbf{Delay level \((\ell)\)} & \(N_{\rm tot}\) & \(N_{\rm act}\) & \(N_{\rm skip}\) & \(\eta_{\rm skip}\) [\%] \\
\hline
1 & 189 & 189 & 0 & 0.0 \\
2 & 2,079 & 1,765 & 314 & 15.1 \\
3 & 27,216 & 20,160 & 7,056 & 25.9 \\
4 & 303,552 & 206,579 & 96,973 & 31.9 \\
5 & 1,728,000 & 1,079,976 & 648,024 & 37.5 \\
6 & 12,856,320 & 7,750,147 & 5,106,173 & 39.7 \\
7 & 99,532,800 & 58,446,293 & 41,086,507 & 41.3 \\
8 & 5,958,400 & 3,963,327 & 1,995,073 & 33.5 \\
\hline
\end{tabular}
\end{table}

Therefore, the interaction counts reported in Table~\ref{tab:causality_statistics} correspond to unique displacement--delay entries of the retarded dyadic Green--function interaction-matrix assembly, including the retained dyadic components, rather than all repeated voxel-pair matrix entries. Here, \(k\) denotes the MOT temporal delay index, whereas \(\ell\) denotes the delay level used to organize the Green--function interaction-matrix assembly.
Table~\ref{tab:causality_statistics}(a) summarizes the resulting interaction reduction. Causal-support screening eliminated 40.6\% of the candidate interactions prior to Green-function evaluation. The per-level statistics in Table~\ref{tab:causality_statistics}(b) shows that causality-based pruning becomes more effective at larger delay levels, with the skipped-interaction ratio increasing from \(0\%\) at \(\ell=1\) to \(41.3\%\) at \(\ell=7\). Physically, larger \(\ell\) represent longer propagation delays, for which more candidate source--observer pairs lie inside the admissible causal support as shown in Figure~\ref{fig:causality_aware_concept}. The decrease at \(\ell=8\) arises from end of the computational domain, where fewer admissible interactions remain admissible.

\begin{table}[htb]
\caption{Vectorized batch-size study using \(P=64\) workers.}
\label{tab:batch_size_study}
\centering
\scriptsize
\setlength{\tabcolsep}{2.8pt}
\renewcommand{\arraystretch}{1}
\begin{tabular}{lccccccc}
\hline
\hline
\(\mathcal{B}_b\) & 500 & 1000 & 2000 & 4000 & 6000 & 8000 & 10000 \\
\hline
\(T_P\) [s] & 2901 & 1709 & 1206 & 1057 & 1047 & \textbf{984} & 1171 \\
Mem. [GB] & 120.0 & 124.8 & 127.8 & 127.4 & 127.1 & 131.7 & 138.3 \\
\hline
\hline
\end{tabular}
\end{table}

To evaluate the batching strategy of Section~\ref{subsec:vectorized_batching}, the number of admissible interactions per batch \(\mathcal{B}_b\) is varied using \(P=64\) CPU workers. Table~\ref{tab:batch_size_study} shows that increasing the batch size from 500 to 8000 interactions reduces the assembly time from \(2901~\mathrm{s}\) to \(984~\mathrm{s}\). Further increases degrade performance, likely due to larger temporary-workspace overhead. Therefore, batches containing 8000 admissible interactions are used in the large-slab case. The number of CPU workers is varied from \(P=1\) to \(P=64\). The corresponding speedup $S_P$ and efficiency $\eta_P$ are defined as
\begin{equation}
S_P=\frac{T_1}{T_P},
\qquad
\eta_P=\frac{S_P}{P}\times100\%,
\label{eq:speedup_efficiency}
\end{equation}
where \(T_1\) and \(T_P\) are the causality-aware assembly times using one and \(P\) workers, respectively. Measurements were performed in MATLAB R2023b on a dual-AMD EPYC 7573ZX shared-memory system with \(2~\mathrm{TB}\) memory. Reported runtimes account only for retarded Green-function assembly.

\begin{figure}[htb]
\centering
\includegraphics[width= 0.8\columnwidth]{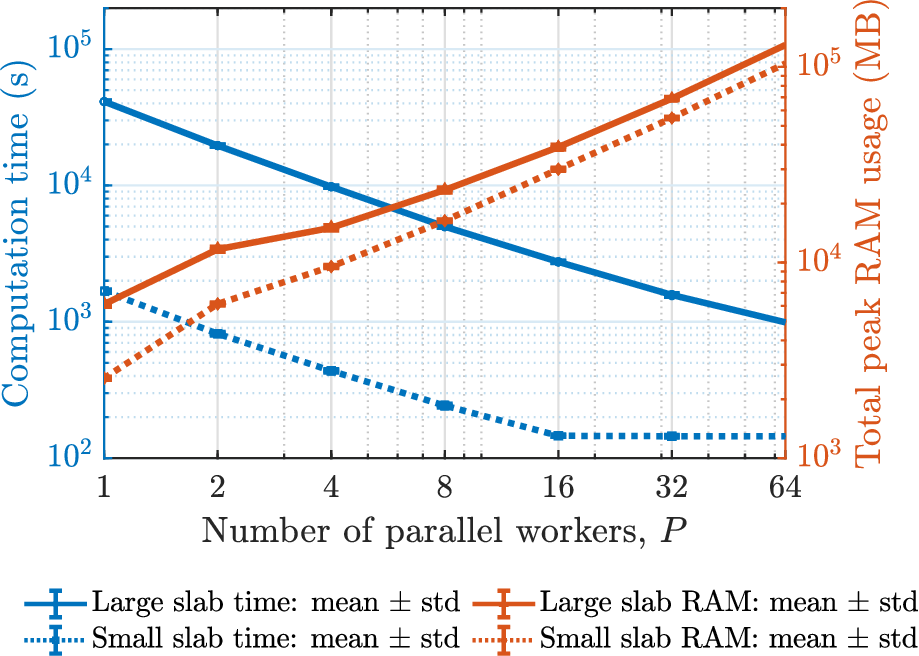}
\caption{CPU-parallel assembly time and peak memory versus the number of workers for small and large slabs. }
\label{fig:time_memory_scaling}
\end{figure}

\begin{figure}[htb]
\centering
\includegraphics[width=0.7\columnwidth]{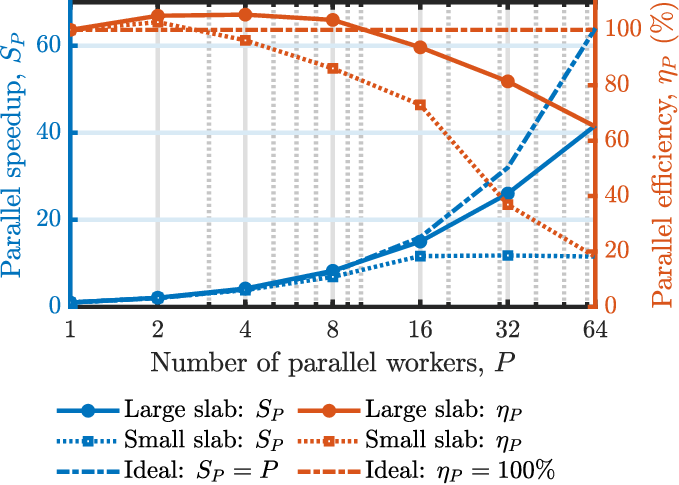}
\caption{Parallel speedup and efficiency of the proposed causality-aware Green-function assembly for the small and large slabs using \(B=8000\).}
\label{fig:speedup_efficiency}
\end{figure}

Figure~\ref{fig:time_memory_scaling} compares the assembly time and peak memory as functions of \(P\) for the small- and large-slab cases. For the large slab, the conventional assembly requires \(1.08\times10^{5}~\mathrm{s}\), whereas the proposed assembly decreases from \(4.1\times10^{4}~\mathrm{s}\) at \(P=1\) to approximately \(990~\mathrm{s}\) at \(P=64\). This corresponds to a \(2.6\) times algorithmic speedup from causal-support screening and a total speedup of \(109\) times relative to the conventional single-worker assembly. Figure~\ref{fig:speedup_efficiency} shows that the proposed assembly achieves a parallel speedup of 42 times at \(P=64\), with efficiency exceeding \(90\%\) up to \(P=16\) and remaining \(65\%\) at \(P=64\). The increase in peak memory with \(P\) is attributed to worker-local interaction batches and accumulation buffers. Overall, causal-aware screening reduces the number of assembled retarded interactions, while vectorized parallel assembly accelerates the remaining admissible interaction history.

\section{Conclusion}
A causality-aware interaction-selection strategy for time-domain integral-equation assembly has been presented. By excluding interactions outside the causal support of the retarded Green function prior to evaluation, the proposed approach reduced the assembly workload by 41\% without approximation and a 2.6 times single-worker algorithmic speedup. Combined with vectorized parallel batching, the proposed approach achieved an acceleration of the assembly up to \(109\) times. The proposed assembly achieved a 42 times parallel speedup with \(65\%\) efficiency at \(P=64\), while peak memory increased approximately linearly with the number of workers due to worker-local batches and accumulation buffers. These results demonstrate that retarded causality can be used as an exact, physics-driven pre-assembly criterion for reducing the cost of large-scale transient electromagnetic scattering simulations.

\bibliographystyle{IEEEtran}
\bibliography{references}

\end{document}